# Significant improvement in estimation of the Young's modulus of the metal foam based on the Timoshenko's bend theory

Tibor Mazúch

*Nábrežná 13/23, SK - 03861 Vrútky, Slovak republic*
*E-mail: tmazuch@gaya.sk*

A computational and experimental approach based on a natural vibration of a free prismatic thick beam with square cross-section is suggested. Three variants of the beam sample were used (one with skin and two without skin). From 12 to 16 lowest resonant frequencies of longitudinal, torsional and flexural vibration of each beam were analyzed. A rule for dependence of Young's modulus on the average foam density was derived from the sample without skin. It is shown that the skin presence causes known anisotropy that the stiffness in transverse direction is about 50% greater than that in the longitudinal direction. It has also been shown that the seeming frequency dependence of the Young's module can be explained by non-uniform distribution of mass density in the sample. Agreement among experimental and numerical data is excellent in most cases. The rule is also verified on solution of two lowest resonant frequencies of the free foam plates with various mass densities (from 400 to 1350 kg/m$^3$). Agreement among experimental and numerical frequencies is acceptable.

## 1. Introduction

One from important factors influencing the expansion of the usage of the aluminium foam in the technical praxis is unquestionably the ability for sufficiently reliably prediction of the behavior of foamed component parts in the operating mode. Besides undisputable interesting properties of this perspective material, is also well known, that the modelling of the response of the metal foam with closed cells on the mechanical forcing is very difficult. Analytical approaches known from courses of the technical (or advanced) mechanics are usable only for few simplest constructional shapes. From several numerical methods the finite element method (FEM) was mostly expanded in this field, see Fig. 1 for an illustration. It is consequence not only of the possibility of modelling complicated shapes and complicated boundary conditions, but also the possibility of the considering various material properties in various positions inside the foam bodies. Reliability of the modelling by the FEM strongly depends on the input data accuracy, including also material constants. For simpler types of forcing and shapes of component parts also approaches based on estimations of moments of inertia are sufficient alternative with respect to finite methods [4-6].

Therefore the aim of this paper is the suggestion and the verification of a reliable approach for the stating Young's modulus in dependence on the foam density. This basic material characteristic is determined from natural vibration of rectangular sample of the foam.

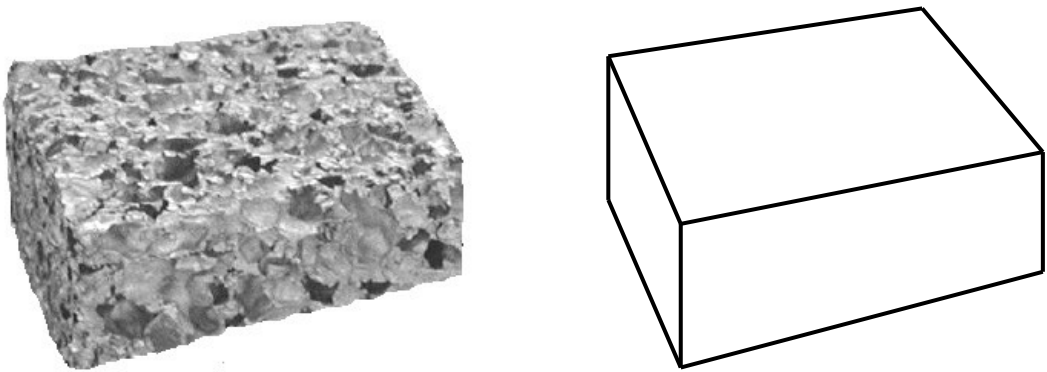

Fig. 1 A real element of the aluminium foam and its FE idealisation

## 2. Suggested approach

From available samples we chose a prismatic beam with square cross-section. Besides good manipulation with the sample during the experiment due to dimensions and the weight (see table 1), is valuable also occurrence of longitudinal, torsional and also flexural mode shapes (see figs. 2 – 4) in chosen frequency range. This allow us multiple verification of results. *Due to square cross-section are all flexural resonant frequencies twice multiple, so frequency spectrum in chosen finite interval is (with respect to the rectangle cross-section) less dense.* Furthermore, possible significant split of twice multiple resonant frequencies can indicate major defects of inner structure of the sample.

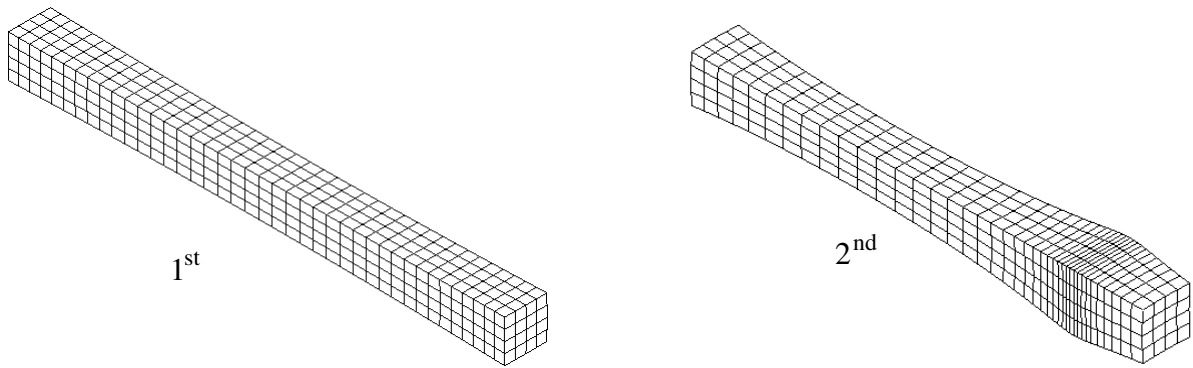

Fig. 2 Longitudinal mode shapes

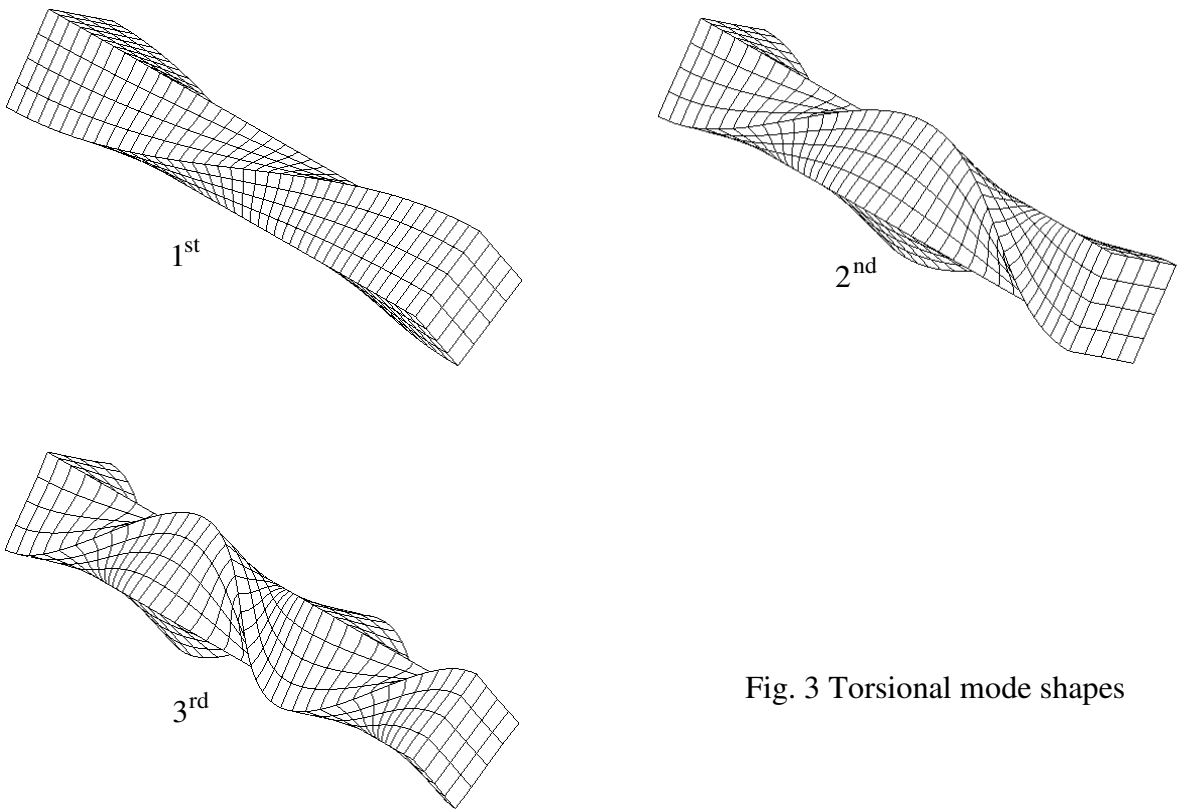

Fig. 3 Torsional mode shapes

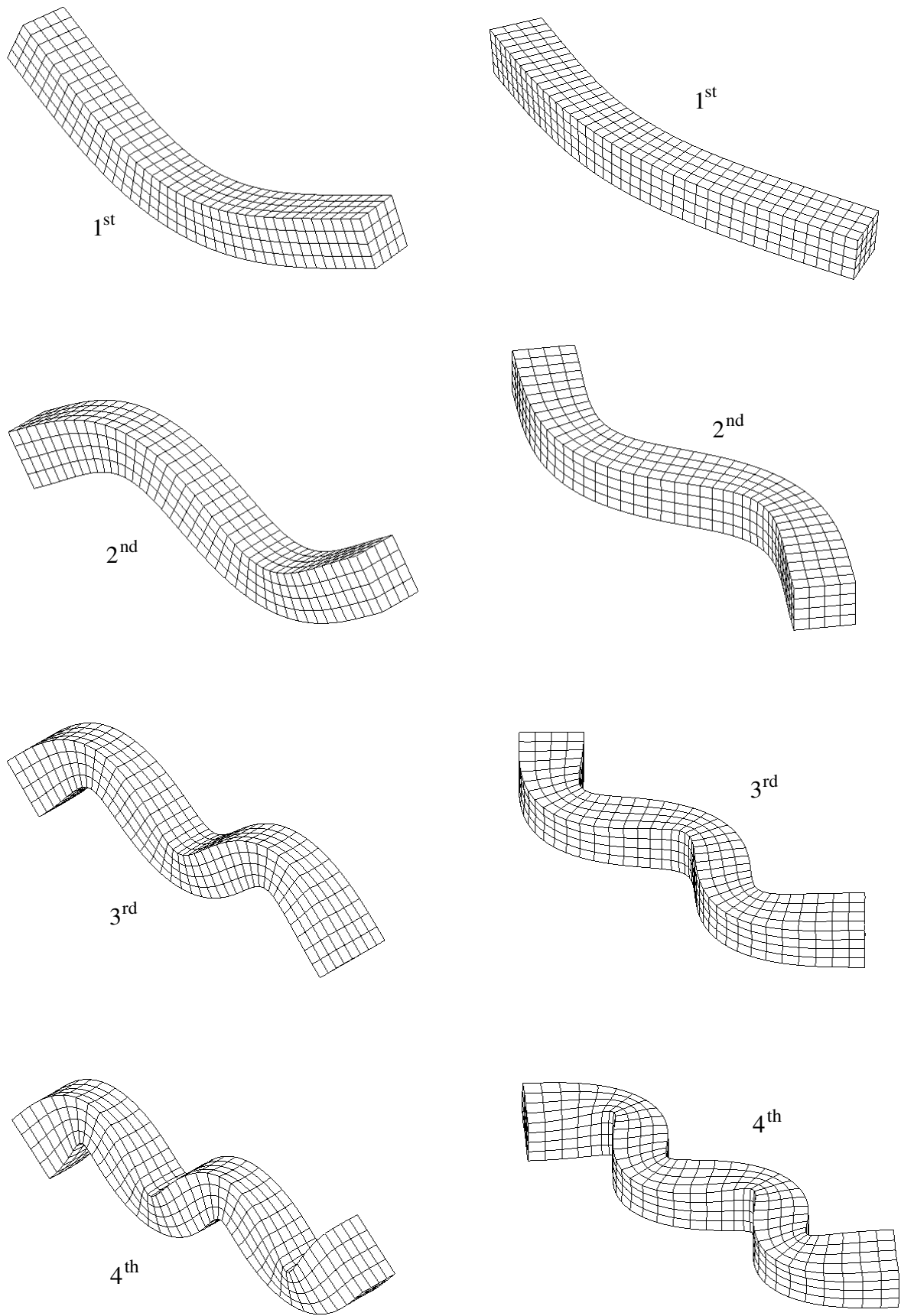

Fig. 4 Flexural mode shapes

Chosen prismatic beam from aluminium foam has following parameters:

Table 1 Basic parameters of three variants of the sample.

| Parameter | | Beam with the skin | Beam without the skin | |
|---|---|---|---|---|
| | | 1st variant | 2nd variant | 3rd variant |
| Cross-section side length | $a_p$ | 49 mm | 46.45mm | 44 mm |
| Beam length | $l$ | 448 mm | 444 mm | 440 mm |
| Beam weight | $m$ | 0.620 kg | 0.4531 kg | 0.3684 kg |
| Average mass density | $\rho$ | 576 kg/m$^3$ | 472 kg/m$^3$ | 432.5 kg/m$^3$ |

The experimental modal analysis was based on the evaluation of the frequency response functions measured by means of a Bruel & Kjaer 2034 FFT Analyzer. Used experimental approaches, including a simulation of boundary conditions, the exciter and the transducer positions, are evident from Figures 5 and 6. The need for a torsional exciter which we have not in disposition, was avoided as follows. With asymmetric excitation and scanning (see Fig. 6), the frequency characteristic for simultaneously excited flexural and torsional vibration was obtained. Then, searched torsional resonant frequencies were identified by using the comparison with characteristic for ‚pure' flexure.

Several configurations of the measurement with various positions of the exciter and the transducer were attempted. The reproducibility of attempts was excellent. The same values of resonant frequencies were mostly obtained. Occasionally the difference between them was equal to the analyzer resolution limit (8Hz) and rarely twice (16 Hz).

Measurements were performed on 3 variants of the sample, see the Table 1. Firstly were measured frequencies on the beam with the skin, then on the beam after 1st machining and finally, on the beam after 2nd machining.

**To avoid a significant distortion of the results in determining the Young's modulus, it was necessary use experimental results of the sample without skin**. The 2nd variant was used for the reason, because for this configuration all 11 theoretical lowest resonant frequencies (2 longitudinal, 4 torsional and 5 flexural) were excited and identified at the chosen interval (0, 6400 Hz). It was mentioned, that all resonant frequencies of the flexural vibrations are twice multiple with independent natural modes (in two mutually perpendicular planes). So, finally, 16 lowest frequencies were measured for that configuration.

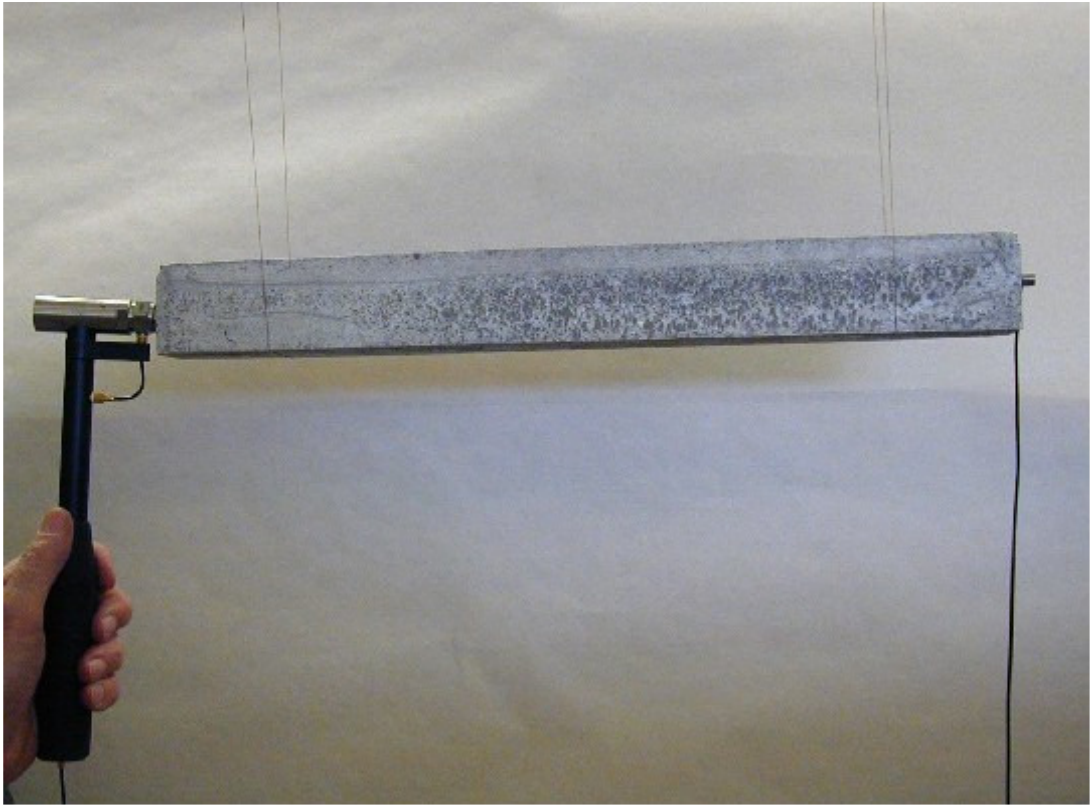

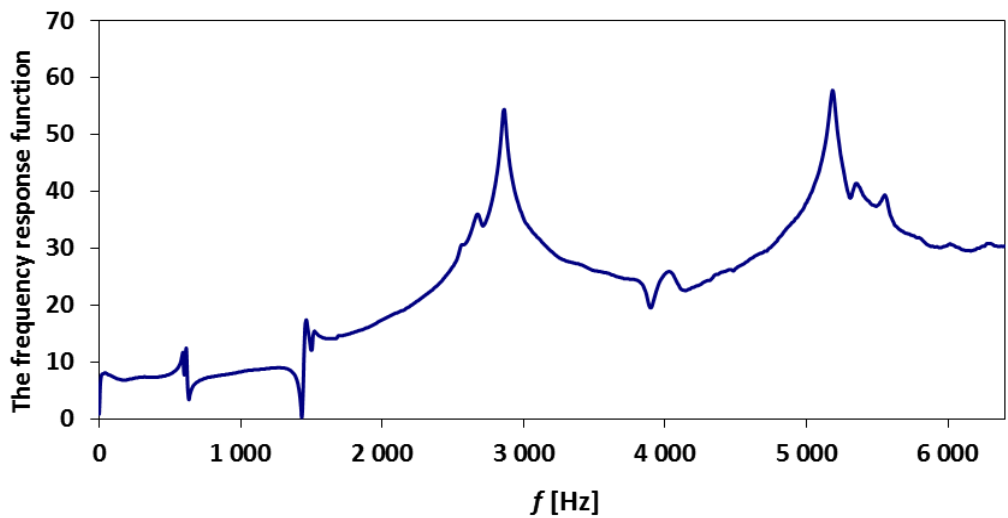

Fig. 5 Experimental set up to obtain longitudinal resonant frequencies

LEGEND:   Up – Simulation of free boundary conditions, and exciter vs transducer positions.

Down – The frequency response function for obtaining first two longitudinal natural frequencies.

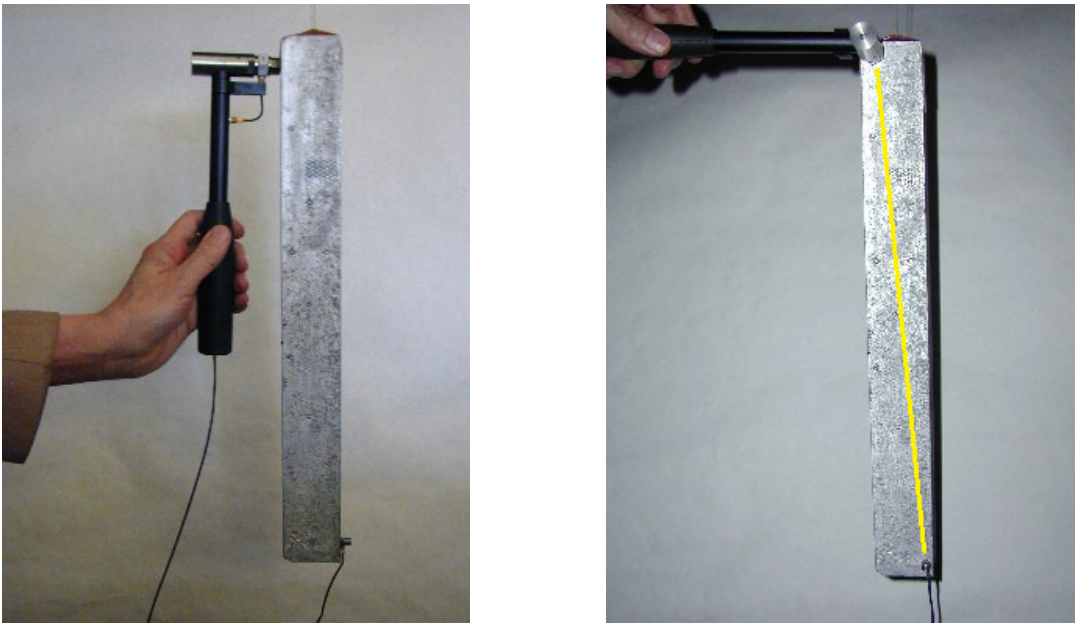

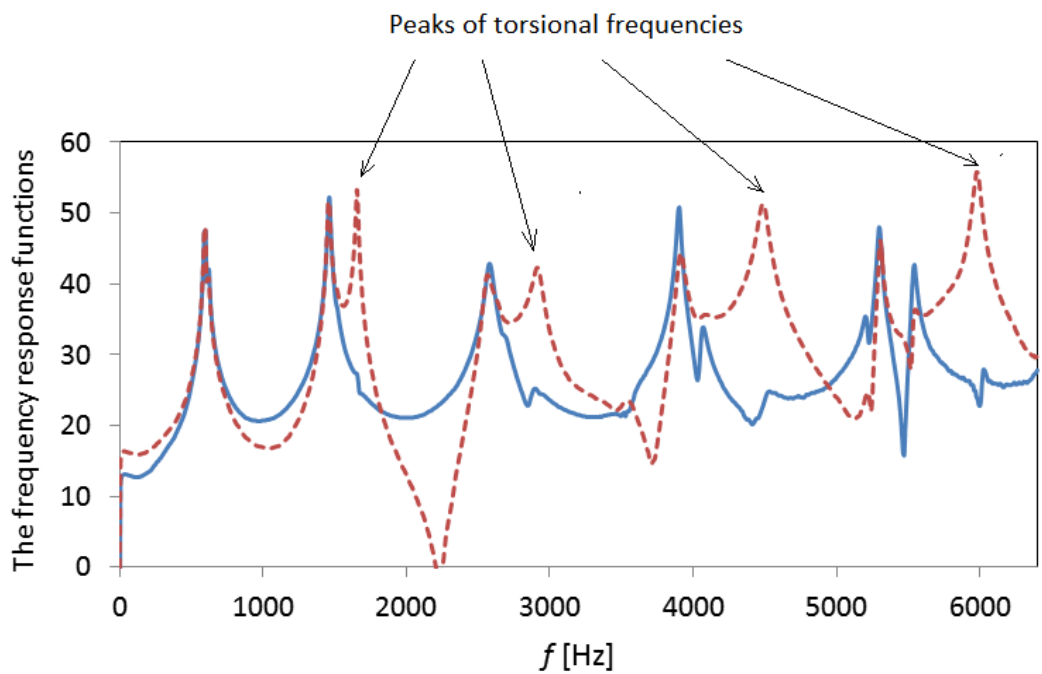

Fig. 6 Experimental set up to obtain torsional and flexural resonant frequencies

LEGEND: Up – Simulation of free boundary conditions, and exciter vs transducer positions.
- Left – for the excitation of flexural mode shapes, only.
- Right – asymmetric (diagonal set up) for simultaneous excitation of torsion and bending shapes.

Down – The frequency response functions (FRP)
- Full line – FRP of flexural vibration, only.
- Dashed line – FRP of simultaneous torsional and flexural vibration.

## 3. Young's modulus in dependence on the mass density

Based on the hypothesis about the quadratic law of the dependence of the foam Young's modulus on the foam mass density was derived several relationships between them. Tobolka and Kováčik derived in [2] a relation from 1st (or lowest) eigenfrequency of free square plate

$$E_{1,T} = 7.38 \cdot 10^3 \cdot \rho^2 + 2.86 \cdot 10^6 \cdot \rho + 2.77 \cdot 10^9, \tag{1a}$$

which was used for the estimation of Young's modulus for described beam with the skin in [1]. Based on its use an excellent approximation of the 1st and very good approximation of the 2nd resonant frequency of the flexural vibration with respect to the experimental results were obtained. But this relation has also certain disadvantage. For the case of $\rho = 0$ (zero weight), nonzero and furthermore relatively great Young's modulus $E = 2.77$ GPa is obtained. It restricts the equation validity for the foam with lower density. It should be noted that from 2nd eigenfrequency free square plate Tobolka and Kováčik derived another equation

$$E_{2,T} = 7.70 \cdot 10^3 \cdot \rho^2 + 2.00 \cdot 10^6 \cdot \rho + 2.65 \cdot 10^9, \tag{1b}$$

which gives less estimation of the Young's modulus for the same mass density $\rho$ …

Fleck, Ashby and co-workers suggested in [3] another equation

$$E/E_0 = 0 \cdot 3\Phi^2(\rho/\rho_0)^2 + (1-\Phi)(\rho/\rho_0); \Phi = 0.7, \tag{2}$$

where $E_0$ and $\rho_0$ are constants of the pure aluminium. Hence

$$E/E_0 = 0 \cdot 147(\rho/\rho_0)^2 + 0 \cdot 3(\rho/\rho_0). \tag{3}$$

For the case of pure aluminium (not foam), i.e. $\rho/\rho_0 = 1$ we obtain less than half of the value of the Young's modulus for aluminium $E/E_0 = 0 \cdot 447$. It restricts the equation validity for the foam with higher density.

It is evident, that for satisfaction of the condition $E(\rho = 0) = 0$ the quadratic equation can sustain from quadratic and linear member, only.

$$E(\rho) = a\rho^2 + b\rho \tag{4}$$

In that case for the unequivocal derivation of the function $E = E(\rho)$ will suffice reliably stated Young's modulus only for one value of the foam density $E_1 = E_1(\rho_1)$. Needed coefficients *a* and *b* by solution of an equation

$$\begin{bmatrix} \rho_0^2 & \rho_0 \\ \rho_1^2 & \rho_1 \end{bmatrix} \begin{Bmatrix} a \\ b \end{Bmatrix} = \begin{Bmatrix} E_0 \\ E_1 \end{Bmatrix}, \tag{5}$$

can be obtained, where $E_0$ and $\rho_0$ are values of the pure aluminium, i.e $E_0$=69 GPa and $\rho_0$=2700 kg/m$^3$. Then

$$\begin{Bmatrix} a \\ b \end{Bmatrix} = \begin{bmatrix} \rho_0^2 & \rho_0 \\ \rho_1^2 & \rho_1 \end{bmatrix}^{-1} \begin{Bmatrix} E_0 \\ E_1 \end{Bmatrix}. \quad (6)$$

From the basic course of the strength theory for the case of linear, isotropic and elastic material is known a relation

$$G = \frac{E}{2(1+\nu)} \quad (7)$$

or

$$E = 2G(1+\nu), \quad (8)$$

where $G$ is the shear modulus and $\nu$ is the Poisson's ratio. Then holds

$$\nu = \frac{E}{2G} - 1. \quad (9)$$

Table 2 Computed material characteristics from measured resonant frequencies sample without the skin with mass density $\rho = 472$ kg/m$^3$

| Mode shape description | Experiment $f$ [Hz] | Shear modulus $G$ [GPa] | Young's modulus $E$ [GPa] |
|---|---|---|---|
| **1st longitudinal** | **2976** | | **3.237** |
| 2nd longitudinal | 5376 | | 2.641 |
| **1st torsional** | **1716** | **1.276** | **3.368** |
| 2nd torsional | 3036 | 1.017 | 2.641 |
| 3rd torsional | 4680 | 1.074 | 2.792 |
| 4th torsional | 6184 | 1.055 | 2.792 |
| **1st flexural** | **637** | | **3.521** |
| **1st flexural** | **662** | | **3.803** |
| 2nd flexural | 1549 | | 3.105 |
| 2nd flexural | 1549 | | 3.105 |
| 3rd flexural | 2720 | | 2.904 |
| 3rd flexural | 2720 | | 2.904 |
| 4th flexural | 4130 | | 2.792 |
| 4th flexural | 4130 | | 2.792 |
| 5th flexural | 5552 | | 2.544 |
| 5th flexural | 5552 | | 2.544 |

The shear modules from resonant frequencies of torsional mode shapes were calculated by the relation (A-4) (see Appendix). Resonant frequencies of longitudinal mode shapes by analytical relation (A-1) do not depend on the Poisson's ratio $\nu$. Based on 3D FE modeling this experience was verified and negligible influence was finded in the study [1]. Carefully we investigated the influence of Poisson's number on the accuracy of the Young's modulus calculated from the flexural resonance frequency, see (A-7) in the Appendix. We have calculated the modules from the 5 lowest frequencies for Poisson numbers from the interval $\nu \in \langle 0.29; 0.34\rangle$. The largest relative difference occurred at 5th frequency and was less than 1.23%. The largest relative difference at the 4$^{th}$ frequency was less than 0.93 %, etc. So we can say that the influence of Poisson's number on the precision of determination of the Young's module is negligible.

Ashby et al., in their monographs [9], indicate the range of Poisson's number from 0.31 to 0.34 for all commercial aluminium foams. Therefore, we have decided to use the value for pure aluminum $\nu$=0.32, which is almost at the center of the mentioned interval.

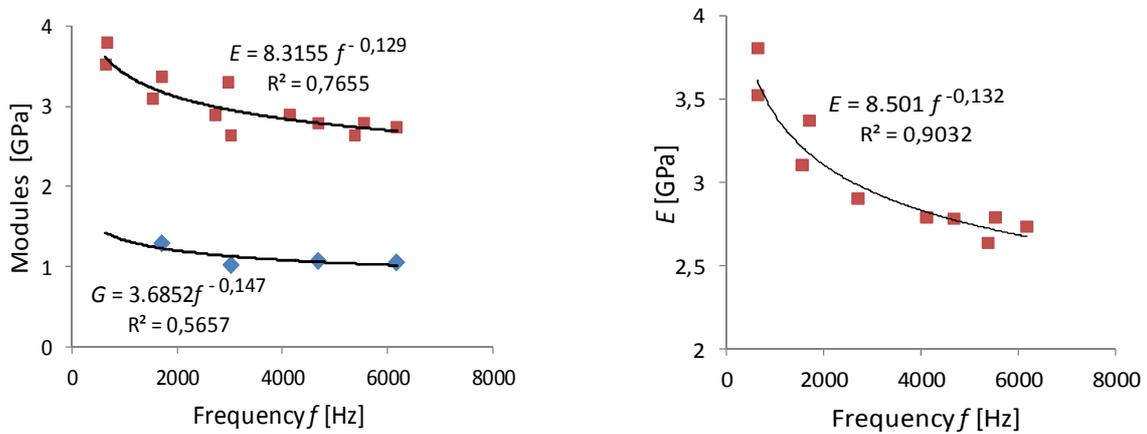

Fig. 7 The apparent frequency dependence of the sample modules

The Table 2 shows an acceptable agreement among Young's modules calculated from resonant frequencies corresponding with simplest mode shapes of longitudinal, torsional and flexural vibration (with except of the 2$^{nd}$ copy of the 1$^{st}$ flexural mode shape). However they are higher than modules calculated from frequencies, which correspond with higher mode shapes. *It looks like the foam Young's modulus, is frequency-dependent. It decreases, when frequency of oscillation increases, so it is not constant. Reality is different. This phenomenon could be explained by non-uniform distribution of the sample density.* Consider, for example, the largest deformations in 1$^{st}$ bending mode shapes, see Fig. 4. If the density of the sample is greater in the central region then logically there is also a greater bending stiffness (resistance against bending deformations). Largest deformations in next bending modes are in places with less density. We will discuss the problem in more detail when verifying the proposed procedure on other experimental data.

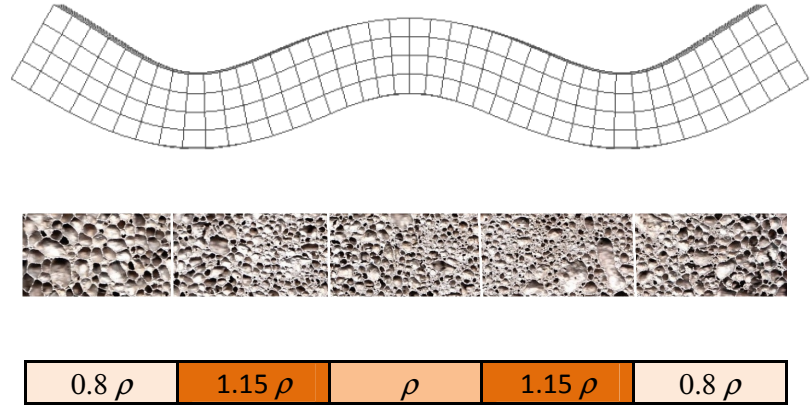

Fig. 8 Roughly estimated density distribution in the sample for anisotropic model

Furthermore, it is known fact that boundary conditions (in the experiment they are not totally equal with conditions considered in calculations) most considerable affect those resonant frequencies, which correspond with the simplest mode shapes (1st in this case). Therefore it is probable that more realistic will be coefficients obtained from frequencies which correspond with all measured resonant frequencies.

Three values of the foam Young's modulus $E_1$ for obtaining coefficients of the equation (4) were considered:

a) $E_{1,a}$ = 3.51 GPa – average value of Young's modules calculated from resonant frequencies corresponding with the simplest (1st) mode shapes (highlighted in the Table 2). It should be noted that most of approaches for obtaining the function $E=E(\rho)$ is based on measuring the lowest frequencies of the samples. Then

$$E_{1,a}(\rho) = 10^6 (0.00813\rho^2 + 3.60\rho). \tag{10}$$

b) $E_{1,b}$ = 3.028 GPa – average value of Young's modules calculated from all measured resonant frequencies

$$E_{1,b}(\rho) = 10^6 (0.00859\rho^2 + 2.36\rho) \tag{11a}$$

or in dimesionless form

$$E_{1,b} / E_0 = 0.908(\rho/\rho_0)^2 + 0.092(\rho/\rho_0), \tag{11b}$$

where $E_0$ and $\rho_0$ are constants of the pure aluminium.

c) Only for an illustration we introduce next equation obtained by using measured resonant frequencies corresponding with the 1st (or the simplest) mode shapes of the flexural vibration of the sample with the skin (1st configuration from the table 1). Then for the average value $E_{1,c}$ = 6.052 GPa

$$E_{1,c}(\rho) = 10^6 (0.00645\rho^2 + 8.14\rho) \cdot \qquad (12)$$

Graphical interpretations of relations (1a), (10-12) are in Fig. 9.

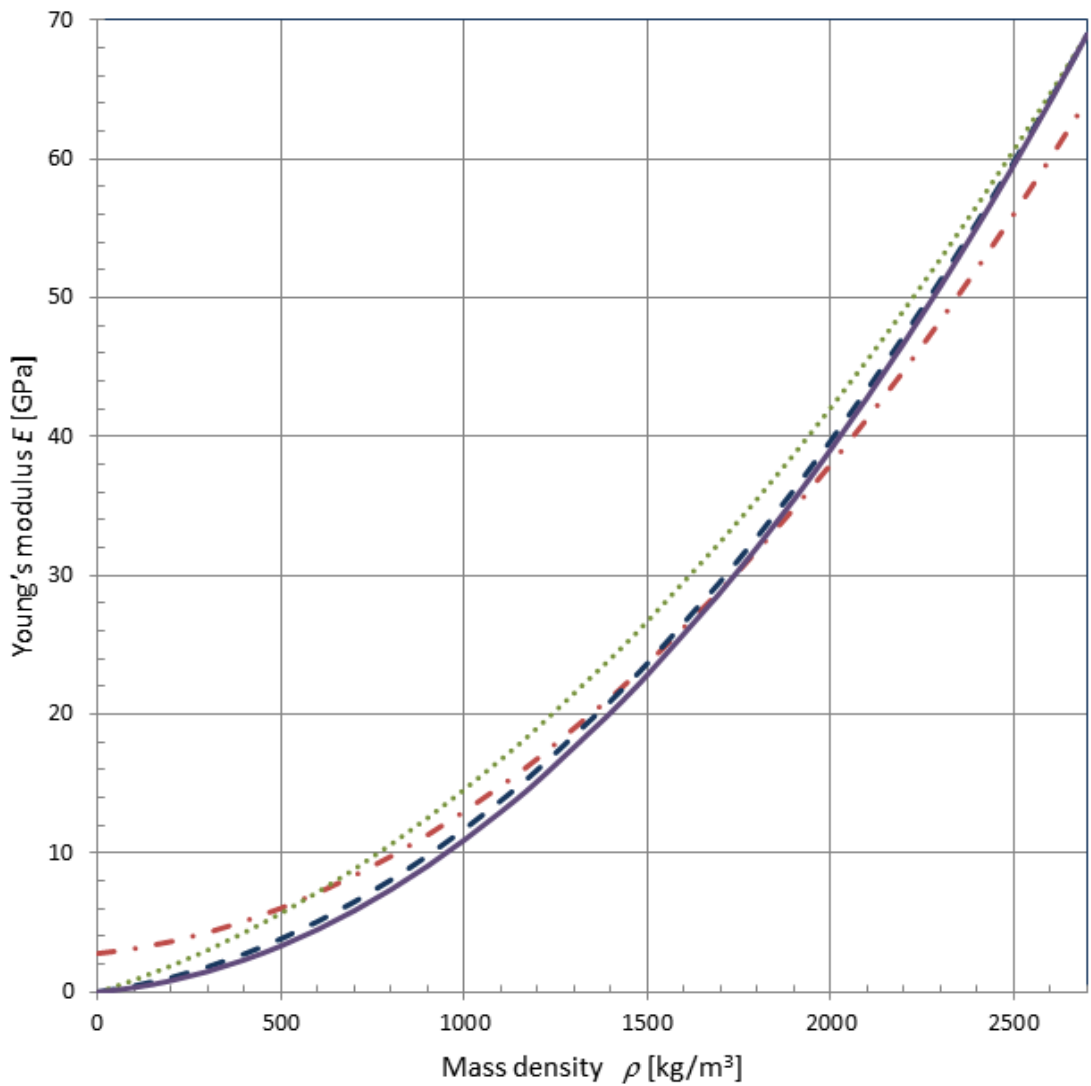

Fig. 9 Dependence of the Young's modulus on the foam mass density

LEGEND: Full line – **recommended values,** by equation (11),
dashed line – values by equation (10),
dotted line – values by equation (12) and
dot-dashed line – values by equation (1a).

## 4. Verification of the approach on experimental results of beams without skin

Derived relationships $E=E(\rho)$ can be verified by using the comparison calculated and experimental resonant frequencies. Consider now the sample with the mass density $\rho = 432.5$ kg/m$^3$ (see the Table 1). Computed mode shapes are in Figures 2 ÷ 4. Results are in the Table 3. It should be noted that **all measured frequencies of bending oscillation are split here**, but the maximum difference between two split frequencies is about 5%. All here used FEM models consisted from quadratic 3D finite elements of serendipity type.

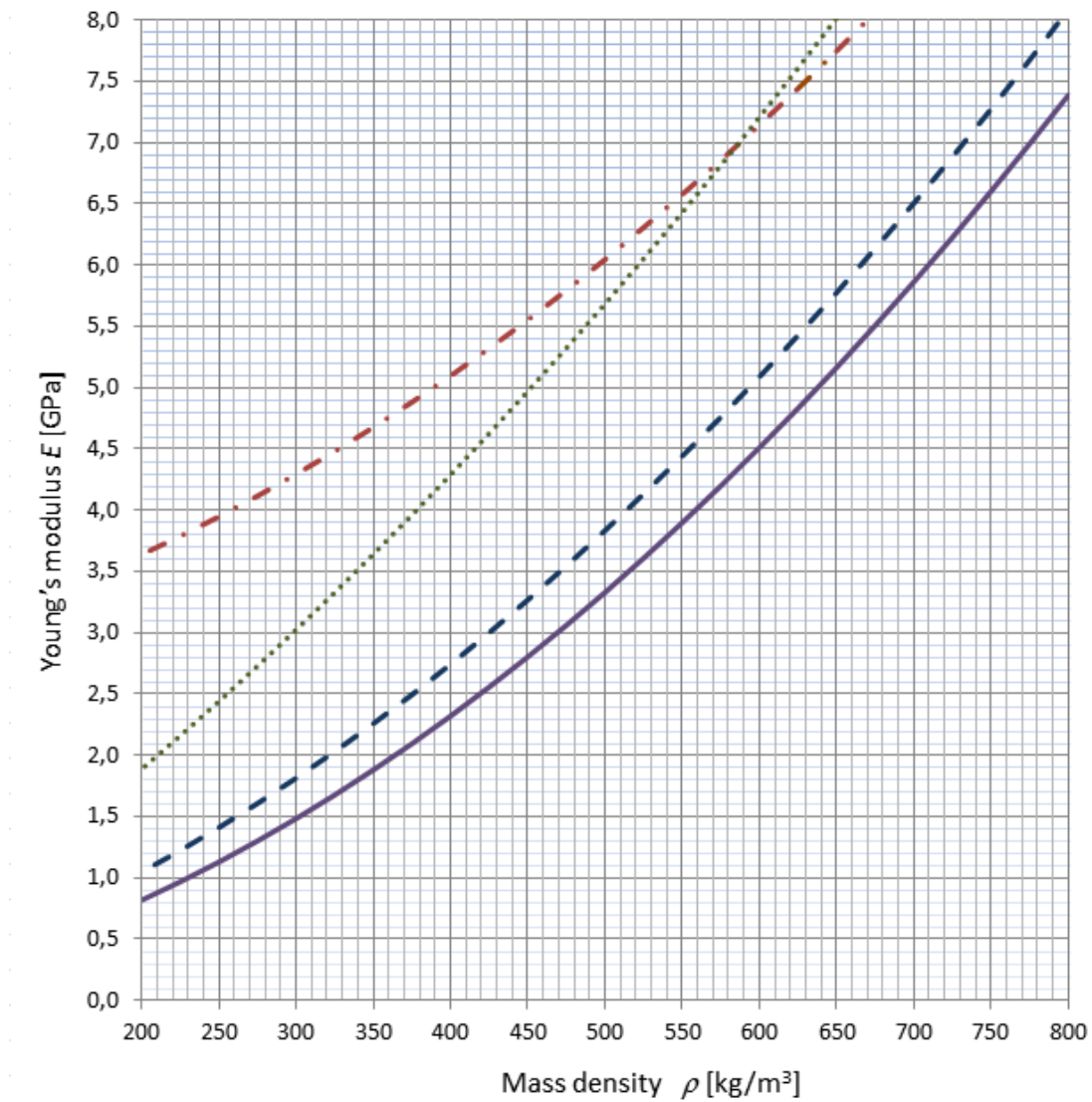

**Fig. 10 Detail from the figure 9 for quick estimation of Young's modulus of the beam without the skin**

LEGEND: Full line – **recommended values,** by equation (11),
dashed line – values by equation (10),
dotted line – values by equation (12) and
dot-dashed line – values by equation (1a).

As expected, only 1$^{st}$ flexural, longitudinal and torsional FE frequencies good approximate experimental results for *E*=3.14 GPa, which was estimated by using eqn. (10). **The other approximations** (except of 2$^{nd}$ copy of 2$^{nd}$ flexural frequency) **are poor**. Better agreement among FE and experimental results was reached for the recommended value *E*=2.63 GPa given by eqn. (11), see also the solid line in Figs 9 and 10.

Table 3 Measured and computed resonant frequencies of the free beam without skin with $\rho = 432.5$ kg/m$^3$

| Mode number | Mode shape description | Experiment Giba | FEM Isotropic model with *E*=3.14 GPa | | FEM Isotropic model with *E*=2.628 GPa | | FEM Anisotropic model with $\rho$ distribution by Fig. 8 | |
|---|---|---|---|---|---|---|---|---|
| | | $f_{g3}$ [Hz] | $f_{m31}$ [Hz] | $\varepsilon_{m31}$ [%] | $f_{m32}$ [Hz] | $\varepsilon_{m32}$ [%] | $f_{m33}$ [Hz] | $\varepsilon_{m33}$ [%] |
| 1 ÷ 6 | as rigid body | ----- | 0 | | 0 | | 0 | |
| 7 | 1$^{st}$ flexural | **592** | 608 | 2.7 | 557 | 5.9 | 608 | 2.7 |
| 8 | 1$^{st}$ flexural | **624** | 608 | 2.5 | 557 | 10.7 | 608 | 2.6 |
| 9 | 2$^{nd}$ flexural | **1432** | 1584 | 10.6 | 1451 | 1.3 | 1469 | 2.6 |
| 10 | 2$^{nd}$ flexural | **1496** | 1584 | 5.9 | 1451 | 3.0 | 1469 | 1.8 |
| 11 | 1$^{st}$ torsional | **1672** | 1732 | 3.6 | 1586 | 5.1 | 1746 | 4.4 |
| 12 | 3$^{rd}$ flexural | **2528** | 2893 | 14.4 | 2650 | 4.8 | 2621 | 3.7 |
| 13 | 3$^{rd}$ flexural | **2680** | 2893 | 7.9 | 2650 | 1.1 | 2621 | 2.2 |
| 14 | 1$^{st}$ longitudinal | **2952** | 3059 | 3.6 | 2803 | 5.1 | 3085 | 4.5 |
| 15 | 2$^{nd}$ torsional | ----- | 3463 | | 3173 | | 3104 | |
| 16 | 4$^{th}$ flexural | **3968** | 4420 | 11.4 | 4049 | 2.1 | 3980 | 0.3 |
| 17 | 4$^{th}$ flexural | **4008** | 4420 | 10.3 | 4049 | 1.0 | 3980 | 0.7 |
| 18 | 3$^{rd}$ torsional | **4568** | 5195 | 13.7 | 4759 | 4.2 | 4646 | 1.7 |
| 19 | 2$^{nd}$ longitudinal | **5296** | 6103 | 15.2 | 5590 | 5.6 | 5471 | 3.3 |
| 20 | 4$^{th}$ torsional | **6056** | 6926 | 14.4 | 6344 | 4.8 | 6036 | 0.3 |
| Average error | | | | 9.0 | | 4.2 | | 2.4 |

Used anisotropic model was composed from five isotropic regions depicted in Fig. 8. For the density values from the figure, Young's modules were calculated according to Equation (11), see also solid line in Fig. 10. The average error value 2.4 % indicates that this model provided the best results. A similar situation also occurred with a higher density sample, see Table 4.

  Comparison of experimental resonance frequencies of flexural oscillation with the calculated frequencies from the anisotropic FEM model is shown also in Fig. 11. The calculated frequencies for the less dense sample are found between the split experimental frequencies. Their trend lines are almost covered. The frequencies calculated for the denser sample are in very good agreements with experimental ones. These agreements indicate that the chosen anisotropic model very well reflects the reality. In other words, this confirms our hypothesis that there is no frequency dependence of the foam modules, but there is an uneven distribution of foam density in the sample.

Table 4 Measured and computed resonant frequencies of the free beam without skin with $\rho = 472 \text{kg/m}^3$

| Mode number | Mode shape description | Experiment Giba | FEM Isotropic model with $E$=3.513 GPa | | FEM Isotropic model with $E$=3.028 GPa | | FEM Anisotropic model with $\rho$ distribution by Fig. 8 | |
|---|---|---|---|---|---|---|---|---|
| | | $f_{g2}$ [Hz] | $f_{m21}$ [Hz] | $\varepsilon_{m21}$ [%] | $f_{m22}$ [Hz] | $\varepsilon_{m22}$ [%] | $f_{m23}$ [Hz] | $\varepsilon_{m32}$ [%] |
| 1 ÷ 6 | as rigid body | ----- | 0 | | 0 | | 0 | |
| 7 | 1st flexural | **637** | 637 | 0.1 | 590 | 7.3 | 635 | 0.3 |
| 8 | 1st flexural | **662** | 637 | 3.8 | 590 | 10.8 | 635 | 4.0 |
| 9 | 2nd flexural | **1549** | 1650 | 6.5 | 1530 | 1.2 | 1552 | 0.2 |
| 10 | 2nd flexural | **1549** | 1650 | 6.5 | 1530 | 1.2 | 1552 | 0.2 |
| 11 | 1st torsional | **1716** | 1738 | 1.3 | 1611 | 6.1 | 1719 | 0.1 |
| 12 | 3rd flexural | **2720** | 3000 | 10.3 | 2782 | 2.3 | 2761 | 1.5 |
| 13 | 3rd flexural | **2720** | 3000 | 10.3 | 2782 | 2.3 | 2761 | 1.5 |
| 14 | 1st longitudinal | **2976** | 3070 | 3.1 | 2847 | 4.3 | 3055 | 2.7 |
| 15 | 2nd torsional | **3036** | 3475 | 14.5 | 3223 | 6.2 | 3138 | 3.3 |
| 16 | 4th flexural | **4130** | 4563 | 10.5 | 4232 | 2.5 | 4185 | 1.3 |
| 17 | 4th flexural | **4130** | 4563 | 10.5 | 4232 | 2.5 | 4185 | 1.3 |
| 18 | 3rd torsional | **4680** | 5213 | 11.4 | 4834 | 3.3 | 4703 | 0.5 |
| 19 | 2nd longitudinal | **5376** | 6122 | 13.9 | 5677 | 5.6 | 5495 | 2.2 |
| 20 | 5th flexural | **5552** | 6259 | 12.7 | 5804 | 4.5 | 5723 | 3.1 |
| 21 | 5th flexural | **5552** | 6259 | 12.7 | 5804 | 4.5 | 5723 | 3.1 |
| 22 | 4th torsional | **6184** | 6949 | 12.4 | 6445 | 4.2 | 6142 | 0.7 |
| Average error | | | | 8.8 | | 3.8 | | 1.6 |

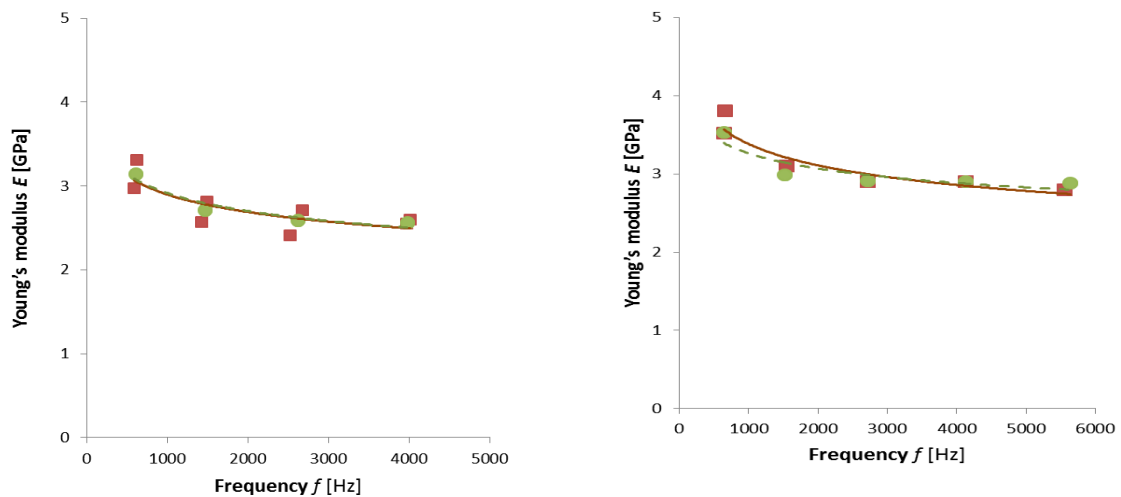

Fig. 11 Anisotropic FEM vs experimental flexural frequencies of beams without skin

LEGEND: Left – sample with $\rho = 432.5 \text{ kg/m}^3$; Right – sample with $\rho = 472 \text{kg/m}^3$;
■ – experimental frequencies; ● - numerical frequencies
full trend line – experimental frequencies; dashed trend line – numerical frequencies

## 5. Verification of the approach on experimental results of the beam with skin

It can be assumed, that the skin of the foam sample is also the foam, but with much higher density. Therefore, for next verification of derived relations $E=E(\rho)$, the results from the experiment of the beam with the skin need be used. Results are in the Table 5. All here used FEM models consisted from quadratic 3D finite elements of serendipity type.

In the classical approach (without separate model of the skin), for the average mass density $\rho$ = 576 kg/m$^3$ the Young's modulus $E_c$ = 6.83 GPa by the eqn. (12) was estimated.

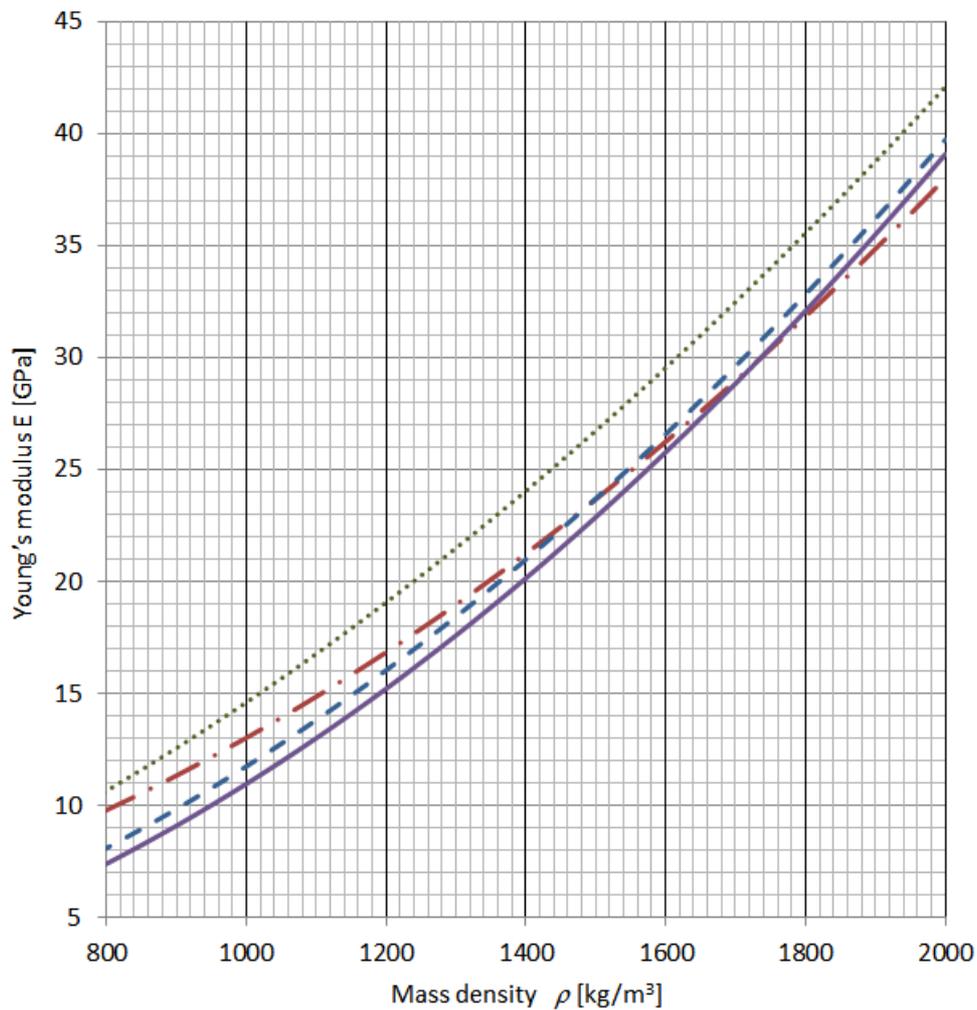

Fig. 12 Detail from the figure 9 for quick estimation of Young's modulus of the foam skin

LEGEND:   Full line – **recommended values,** by equation (11),
dashed line – values by equation (10),
dotted line – values by equation (12) and
dot-dashed line – values by equation (1a).

We noted that Tobolka - Ková čik's relationship (1a) gave almost the same value – $E_{1T}$ = 6.85 GPa (see also the Fig. 10). This totally isotropic model gives good aproximations for both copies of the 1st flexural frequency and the 2nd copy of the 2nd flexural frequency, only. **The other approximations are poor.**

Two variants of anistropic FEM model were considered. The mass density of the skin on opposite square faces was $\rho_{s1}$ = 1692 kg/m³ with $E_{s1}$ = 28.6 GPa. The average mass density of the skin on all other parts of the beam was $\rho_{s2}$ = 1424 kg/m³ with $E_{s2}$ = 20.8 GPa. The inner foam (or the core) of considered sample had average mass density $\rho_f$ = 472 kg/m³ and Young's modulus $E_f$ = 3.03 GPa. It is evident from the table, that this model yields to results in much more greater accuracy than the classical approach.

The 2nd anisotropic model is more complicated. It consists from the core with $\rho$ distribution by the Fig. 8 and two isotropic skins. By other words all anisotropic models used in these analyzes were ,per partes' isotropic. The most accurate results were obtained for the 2nd anisotropic FEM model, see two last columns in Table 5.

Table 5 Measured and computed resonant frequencies of the free beam with the skin

| Mode number | Mode shape description | Experiment Giba $f_{g1}$ [Hz] | FEM Isotropic model with $E$=6.83 GPa $f_{m11}$ [Hz] | $\varepsilon_{m11}$ [%] | FEM Anisotropic model composed from isotropic core and isotropic skin $f_{m12}$ [Hz] | $\varepsilon_{m12}$ [%] | FEM Anisotropic model with considering $\rho$ distribution by Fig. 8 in core and isotropic skin $f_{m13}$ [Hz] | $\varepsilon_{m13}$ [%] |
|---|---|---|---|---|---|---|---|---|
| 1 ÷ 6 | as rigid body | ----- | 0 | | 0 | | 0 | |
| 7 | 1st flexural | **784** | 830 | 5.8 | 799 | 1.9 | 807 | 2.9 |
| 8 | 1st flexural | **856** | 830 | 3.1 | 799 | 6.7 | 807 | 5.7 |
| 9 | 2nd flexural | **1992** | 2141 | 7.5 | 2023 | 1.5 | 1995 | 0.1 |
| 10 | 2nd flexural | **2072** | 2141 | 3.3 | 2023 | 2.4 | 1995 | 3.7 |
| 11 | 1st torsional | ----- | 2174 | | 1953 | | 2032 | |
| 12 | 1st longitudinal | **3295** | 3839 | 16.5 | 3237 | 1.8 | 3406 | 3.4 |
| 13 | 3rd flexural | **3552** | 3873 | 9.0 | 3589 | 1.0 | 3521 | 0.9 |
| 14 | 3rd flexural | **3552** | 3873 | 9.0 | 3589 | 1.0 | 3521 | 0.9 |
| 15 | 2nd torsional | **3768** | 4347 | 15.4 | 3905 | 3.6 | 3787 | 0.5 |
| 16 | 4th flexural | **5176** | 5864 | 13.3 | 5335 | 3.1 | 5239 | 1.2 |
| 17 | 4th flexural | **5176** | 5864 | 13.3 | 5335 | 3.1 | 5239 | 1.2 |
| 18 | 3rd torsional | **5640** | 6520 | 15.6 | 5853 | 3.8 | 5741 | 1.8 |
| 19 | 2nd longitudinal | **6176** | 7655 | 23.9 | 6431 | 4.1 | 6152 | 0.4 |
| Average error | | | | 12.3 | | 3.1 | | 1.9 |

One interesting experience is founded from three presented sets of experimental results (see Tables 5, 4 and 3). It is evident from the Table 6, that for the beam with the skin was calculated from 2$^{nd}$ copy of split 1$^{st}$ flexural resonant frequency the Young's modulus $E$=7.430 GPa and from the 1$^{st}$ longitudinal resonant frequency $E$ = 5.021 GPa. The relative difference between them was about 48 %. Fleck, Ashby with co-workers in [3] described similar experience that strength and stiffness of some foams in the transverse direction is about 50% greater than longitudinal and through thickness directions. They justify it by "shoebox" shaped cells of the foam. After removing of the skin of the sample (for the foam mass density $\rho$ = 472.0 kg/m$^3$) this difference decreased approximately to 15 %. Then after removing next foam layer from all sides of the sample (see last three columns in the Table 5) the difference decreased to 11 %.

Table 6 Foam anisotropy properties of three variants of considered sample

| Mode | Beam with the skin $\rho$ = 576.0 kg/m$^3$ | | | Beam without the skin | | | | | |
|---|---|---|---|---|---|---|---|---|---|
| | | | | $\rho$ = 472.0 kg/m$^3$ | | | $\rho$ = 432.5 kg/m$^3$ | | |
| | $f$ [Hz] | $E$ [GPa] | Difference [%] | $f$ [Hz] | $E$ [GPa] | Difference [%] | $f$ [Hz] | $E$ [GPa] | Difference [%] |
| 1$^{st}$ flexural | 856 | 7.43 | **48** | 662 | 3.80 | **15** | 624 | 3.31 | **11** |
| 1$^{st}$ flexural | 784 | 6.23 | **24** | 637 | 3.52 | **7** | 529 | 2.98 | **1** |
| 1$^{st}$ longitudinal | 3295 | 5.02 | - | 2976 | 3.30 | - | 2952 | 2.96 | - |

It follows that in the case of here-considered sample this strong anisotropy effect was caused above all by the existence of the skin on the sample. Another significant factor was the non-uniform distribution of density in tested sample.

Conformity of numerical calculations with experimental data (see Tables 3, 4 and 5) is limited by the fact that bending frequencies have been split. **This could be improved by considering of so-called bimodular behaviour in transverse directions.** It is possible to simulate this by using a different distribution of density in transversal directions. However, this improvement is not the subject of this publication.

## 6. Verification of the approach on the Tobolka – Kováčik's plates

Here proposed rule to determine the Young's modulus from average foam density (11) we have also tested on the task with a larger range of foam density values.

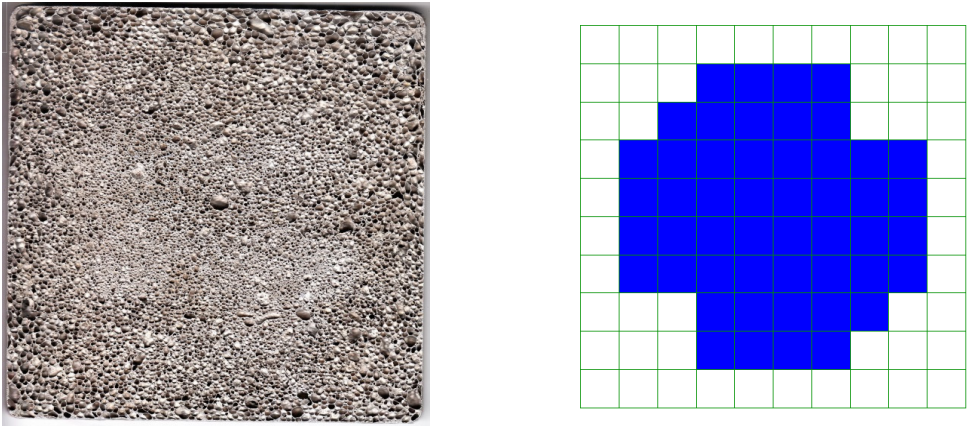

Fig. 13 Tobolka – Kováčik's plate

LEGEND: Left – cross section of the plate
Right – Roughly estimated density distribution in the sample for anisotropic FE model. Highlighted parts have considered density $1.15\rho$ and the other parts have density $0.85\rho$.

Considered plate [2] has dimensions 137 x 137 x 8 mm and for Young's modulus determination was there used the Leissa's relationship [10]. Note that due to the apparent frequency dependence of the module, there were determined 2 different equations to determine the modulus of elasticity in [2], see (1a) and (1b).

$$f_n = \frac{B_n}{2\pi}\sqrt{\frac{Eh^2}{\rho a^4 (1-v^2)}}, \text{ with } B_1 = 4.08 \text{ and } B_2 = 5.91. \qquad (13)$$

By using FEM model (consisted from 3D quadratic elements), we have tested the accuracy of the relationship for the plate from pure iron and aluminium. Error on the 1st frequency was 7.6% and on the 2nd frequency about 6% in both cases. Tobolka – Kováčik's experimental data (not affected by this inaccuracy) can simply be reconstructed by substituting the relation (1a) for the 1st frequency and the relation (1b) for 2nd frequency into (13).

According to the sample from Fig. 13 we have estimated the surface skin thickness to about 0.3 mm on both sides of the plate. The density of the skin ($\rho$ =1424 kg/m$^3$) obtained from previously tested beam was used also in these FEM calculations (for all considered average densities of the plate). Computed mode shapes are on Fig. 14 and the comparison of numerical and experimental resonant frequencies shows Fig. 15.

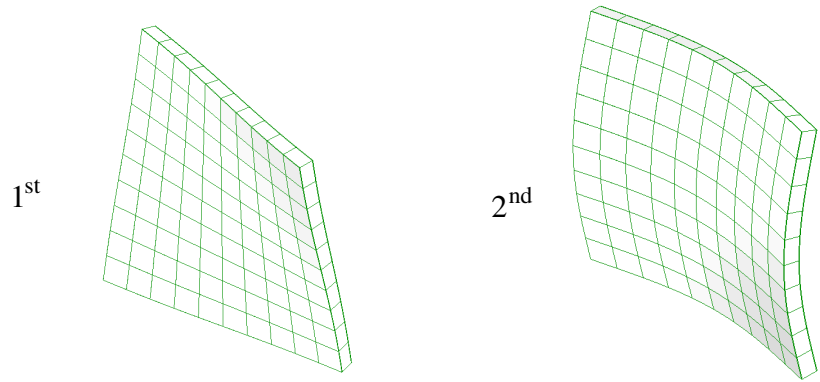

1$^{st}$      2$^{nd}$

Fig. 14 Natural modes corresponding with two lowest resonant frequencies of the plate

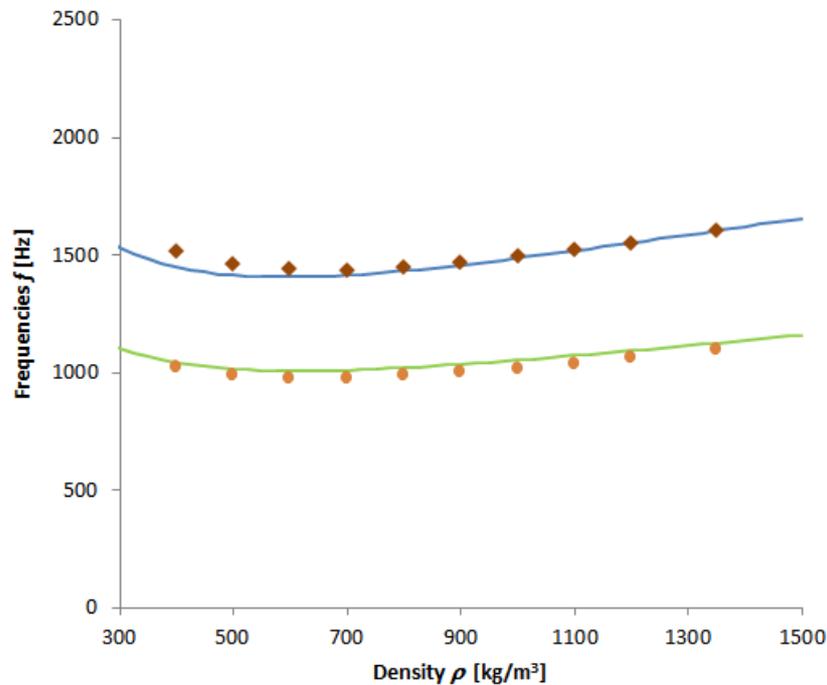

Fig. 15 Comparison of numerical and experimental results on 2 lowest frequencies in dependence on average plate density

    LEGEND:    Full lines **–** Tobolka – Kováčik'experimental data
                             Discrete points **– FEM results with considering only one equation (11) for both lowest resonant frequencies.**

FEM simulations were performed for a plate mass density in the range of 400 to 1350 kg/m$^3$. Largest relative differences between numerical and experimental data less than 3 % on 1$^{st}$ frequency and less than 4.8 % on 2$^{nd}$ frequency were occurred. We noted that very good agreement among these FE results and the experimental data should be treated with some reserve, because the parameters of the skin were only estimated.

## 7. Concluding remarks

Based on the original numerical-experimental way, we obtained a rule for the Young's module's dependence on the density of the aluminum foam, see equations (11a, 11b). The rule was derived from natural vibration of thick prismatic beam with constant square cross-section. Longitudinal, torsional and flexural vibrations were taken into account. The properties of the derived rule are predetermined by the extremely accurate Timoshenko's bend theory and hypothesis about the quadratic law of the module dependence on the foam density.

Validity of here-derived relationship was verified by comparisons of FEM calculations with experimental results on the natural vibration of thick beam. Very good agreements were obtained by using the isotropic FEM models of beams without skin. Anisotropic models (composed of several isotropic regions) gave an excellent agreement with experimental data in most cases. The anisotropic model gave excellent approximations also for the beam with the skin. Finally, we verified the validity of the rule on natural vibration of the free thinner foamed plate with the skin on both sides. The agreement among experimental and numerical data (2 lowest resonant frequencies) of square foamed plates with varying mass densities (from 400 to 1350 $kg/m^3$) was very good, too. *It has been shown that even a rough estimate of density distribution can significantly increase the accuracy of frequency approximations.*

It was demonstrated, that strong anisotropic foam properties were caused mainly by the existence of the skin and non-uniform distribution of the mass density in the foam sample.

The derived relationship (11) could be useful in improving the accuracy of numerical or analytical modeling of mechanical behavior of foam (in various sandwich or functionally graded material approaches, e.g.). The relationship and conclusions are valid for aluminum foams of "closed cells". It would be useful if someone verified the validity of the dimensionless equation (11b) also for other (not aluminum) metal foams.

### Acknowledgement and dedication

**The author would like to dedicate this article to the memory of Dr. V. Giba.** His perfect realization of here proposed experiment is gratefully acknowledged.

# Appendix

## A-1 Nomenclature

| | |
|---|---|
| $E$ | Young's modulus |
| $G$ | shear modulus |
| $\rho$ | mass density |
| $\nu$ | Poisson's number |
| $A$ | area of the cross-section |
| $J$ | area moment of inertia of the cross-section about the neutral axis |
| $J_p$ | polar moment of inertia of the cross-section about the torsional axis |
| $l$ | beam length |
| $k'$ | shape factor of the cross-section |
| $c$ | torsional constant of the cross-section |
| $f_n$ | $n$-th resonant frequency (for all three types of natural vibration) |
| $\overline{\omega}_n$ | $n$-th dimensionless circular frequency (for Timoshenko's theory of natural vibration, only) |
| skin | coversheet of the foam |

## A-2 Used analytical relations for natural vibration of prismatic beam

Here are used relations for resonant frequencies of isotropic free prismatic beam and following relations for Young' modulus.

### *Longitudinal vibration* [8]

By elementary theory for resonant frequencies holds

$$f_n = \frac{n}{2\pi l}\sqrt{\frac{E}{\rho}}, \quad n = 1,2,... \tag{A-1}$$

and hence

$$\boxed{E = \frac{4l^2 f_n^2 \rho}{n^2}}. \tag{A-2}$$

### *Torsional vibration* [8]

For resonant frequencies of the prismatic beam with constant square cross-section holds

$$f_n = \frac{n}{2l}\sqrt{\frac{cG}{\rho J_p}}, \quad \text{where} \quad c = 0.1406\,A, \quad J_p = \frac{A}{6}, \quad n = 1,2,... \tag{A-3}$$

and hence

$$\boxed{G = \frac{(2f_n)^2 l^2 \rho J_p}{cn^2}}. \tag{A-4}$$

***Flexural vibration – Euler-Bernoulli's theory*** [8]

$$f_n = \frac{(\beta l)_n^2}{2\pi}\sqrt{\frac{EJ}{\rho A}}, \quad n = 1,2,... \quad \text{(A-5)}$$

and hence

$$E = \frac{(2\pi f_n)^2 l^4 \rho A}{J(\beta l)_n^4}.$$

For free beam on both ends holds

$$(\beta l)_1 = 4.730, \; (\beta l)_2 = 7.853, (\beta l)_2 = 10.996 \; a \; (\beta l)_n = (2n+1)\frac{\pi}{2}; \; n > 3.$$

***Flexural vibration – Timoshenko's theory*** [5-7]

By Timoshenko's theory for free beam with given geometrical a material properties holds

$$f_n = \frac{1}{2\pi l}\sqrt{\frac{E(a_n^2 - b_n^2)}{\rho(1+\gamma^2)}}, \quad n = 1,2,... \quad \text{(A-6)}$$

where

$$a_n = \sqrt{B_1 + B_2 + \sqrt{(B_1-B_2)^2 + B_3}}, \quad b_n = \sqrt{-B_1 - B_2 + \sqrt{(B_1-B_2)^2 + B_3}},$$

with

$$B_1 = \frac{\rho J \overline{\omega}_n^2}{2}, \; B_2 = B_1 \gamma, \; \gamma = \frac{2(1+\nu)}{k'}, \; B_3 = \frac{\rho A \overline{\omega}_n^2}{L^2}.$$

For a rectangle (and also for a square) holds

$$J = \frac{A^2}{12}, \quad k' = \frac{10(1+\nu)}{12+11\nu}$$

and 5 lowest dimensionless circular frequencies for the beam with the skin and also for two configurations without the skin are in the next table.

Table A-1 Lowest dimensionless circular frequencies of three variants of the foam sample for Poisson's number $\nu = 0.32$

| Configurations of the beam by Table 1 | $\overline{\omega}_1$ | $\overline{\omega}_2$ | $\overline{\omega}_3$ | $\overline{\omega}_4$ | $\overline{\omega}_5$ |
|---|---|---|---|---|---|
| 1st variant | 8.17972 | 21.0813 | 38.0873 | 57.6022 | 78.611 |
| 2nd variant | 9.47838 | 24.5463 | 44.5671 | 67.7106 | 92.7728 |
| 3rd variant | 10.3911 | 27.0339 | 49.3209 | 75.2725 | 103.5430 |

These values were solved from a transcendental equation

$$\frac{(a_n^2-b_n^2)(a_n^2+b_n^2+\gamma^2 a_n b_n - a_n b_n)(a_n^2+b_n^2-\gamma^2 a_n b_n + a_n b_n)}{2a_n b_n (b_n^2+\gamma^2 a_n^2)(a_n^2+\gamma^2 b_n^2)} \sin a_n \sinh b_n - \cos a_n \cosh b_n = -1.$$

Young's modulus can be obtained by

$$\boxed{E = \frac{(2\pi f_n l)^2 \rho(1+\gamma)}{(a_n^2-b_n^2)}}. \qquad (A\text{-}7)$$

It should be noted, that described relations hold for considered configurations of the beam. For the other parameters these can be more complicated.

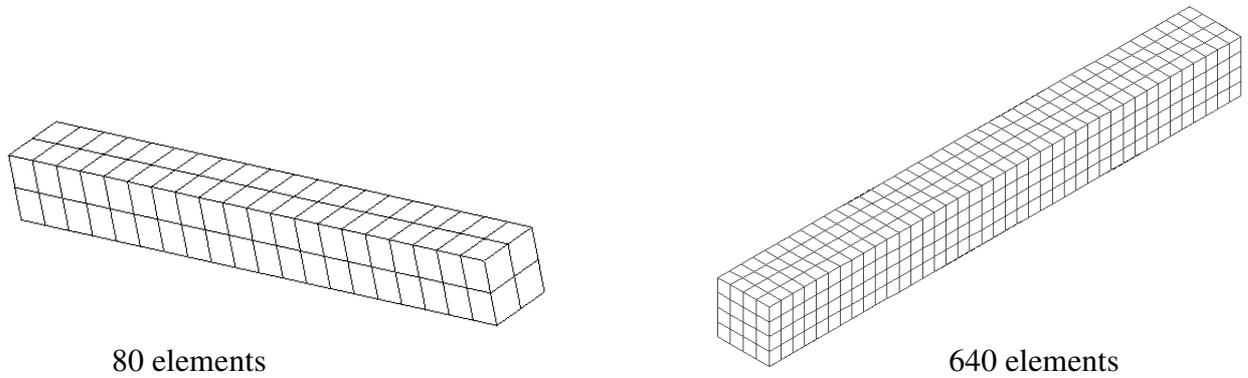

80 elements      640 elements

Fig. A-1 Used FE meshes for the test of covergence

Beam dimensions: 0.049m × 0.049m × 0.448m.

Material constants are taken from [1]:
Young's modulus     $E = 6.85$ GPa
Mass density     $\rho = 576$ kg/m$^3$
Poisson's number     $\nu = 0.3$

In the table A-2 are results from three FE models (composed from 80 to 640 3D quadratic 20-noded solids of serendipity type) and also from analytical theories. It is evident that the FE results are only little varied with respect to the size of the FE model. It follows that, for sufficiently accurate modelling of the problem, even the simplest model is enough.

The agreement of the analytical approach [8] with our FE results for torsional resonant frequencies is excellent. Euler-Bernoulli's (elementary) theory gives acceptable agreement only for the lowest flexural resonant frequency. For higher frequencies it is falling down. On the opposite side the **Timoshenko's theory** [5-7] (results are in the column for engineering theories) **gives an excellent agreement** (relative difference is less than 1%) **also for the 5$^{th}$ resonant frequency**. Agreement among analytical estimations for longitudinal eigenfrequencies with our FE solutions is also excellent (relative difference is less than 0.3%).

### A-3 FEM convergence

Table A-2 Convergence of FE model and the comparison with analytical solutions

| Mode description | FEM 80 elem. $f_{M1}$[Hz] | FEM 270 elem. $f_{M1}$[Hz] | FEM 640 elem. $f_{M2}$[Hz] | **Engineering teories** $f_{et}$[Hz] | Euler-Bernoulli bend theory $f_{eb}$[Hz] |
|---|---|---|---|---|---|
| 6x rigid body mode | 0 | 0 | 0 | | |
| 1$^{st}$ flexural | 833 | 832 | 832 | **831** | 867 |
| 1$^{st}$ flexural | 833 | 832 | 832 | **831** | 867 |
| 2$^{nd}$ flexural | 2151 | 2149 | 2148 | **2142** | 2389 |
| 2$^{nd}$ flexural | 2151 | 2149 | 2148 | **2142** | 2389 |
| 1$^{st}$ torsional | 2212 | 2199 | 2197 | **2195** | |
| 1$^{st}$ longitudinal | 3852 | 3852 | 3851 | **3854** | |
| 3$^{rd}$ flexural | 3895 | 3889 | 3889 | **3872** | 4683 |
| 3$^{rd}$ flexural | 3895 | 3889 | 3889 | **3872** | 4684 |
| 2$^{nd}$ torsional | 4425 | 4389 | 4393 | **4390** | |
| 4$^{th}$ flexural | 5906 | 5893 | 5891 | **5860** | 7742 |
| 4$^{th}$ flexural | 5906 | 5893 | 5891 | **5860** | 7742 |
| 3$^{rd}$ torsional | 6640 | 6597 | 6589 | **6584** | |
| 2$^{nd}$ longitudinal | 7684 | 7681 | 7681 | **7709** | |
| 5$^{th}$ flexural | 8082 | 8057 | 8053 | **8002** | 11565 |
| 5$^{th}$ flexural | 8082 | 8057 | 8053 | **8002** | 11565 |
| 4$^{th}$ torsional | 8857 | 8796 | 8784 | **8780** | |

It can be said, that **analytical theories described in Appendix A-2** (with exception of the Euler-Bernoulli theory) **are sufficiently exact and** therefore **they are sufficient for estimates of material constants** from resonant frequencies of the beam considered.